\documentclass[twocolumn,aps,pra,showpacs]{revtex4}
\usepackage{epsfig}
\begin{document}

\title{To detect the looped Bloch bands of Bose-Einstein condensates
in optical lattices}
\author{Dae-Il Choi}
\affiliation{Universities Space Research Association, 7501 Forbes Boulevard, 
\#206, 
Seabrook, MD 20706 \\ 
Laboratory for High Energy Astrophysics, NASA Goddard Space 
Flight Center, Greenbelt, MD 20771}
\author{Biao Wu}
\affiliation{Department of Physics, The University of Texas, Austin, TX 78712}
\affiliation{Solid State Division, Oak Ridge National Laboratory, Oak Ridge, 
Tennessee 37831-6032}

\date{\today}

\begin{abstract}
A loop structure was predicted to exist in the
Bloch bands of Bose-Einstein condensates in optical lattices
recently in [{\it Wu and Niu, Phys. Rev. A {\bf 61}, 023402 (2000)}].
We discuss how to detect experimentally the looped band with an
accelerating optical lattice through extensive and realistic numerical 
simulations. 
We find that the loop can be detected through observing either nonlinear 
Landau-Zener 
tunneling or destruction of Bloch oscillations. 
\end{abstract}

\maketitle

\section{Introduction}
The simple system of Bose-Einstein condensates (BECs) in 
optical lattices is of amazingly rich physics, as shown in 
recent theoretical and experimental studies. With certain choices 
of densities of BEC and strengths of optical lattice,
this system exhibits various interesting phenomena, ranging
from the dynamics of BEC Bloch waves\cite{choi,wn,zobay,wdn,burger,morsch}, 
Josephson effect\cite{josephson}, squeezed states\cite{squeeze}, 
and quantum phase transition between
superfluidity and Mott-insulator\cite{mott}. 
One can only expect more interesting physics to be discovered
in this system, considering the rich physics that we have known in the
condensed-matter physics, where the prototype system is
electrons in a crystal lattice.

\begin{figure}[!htb]
\centerline{\includegraphics[width=7.5cm]{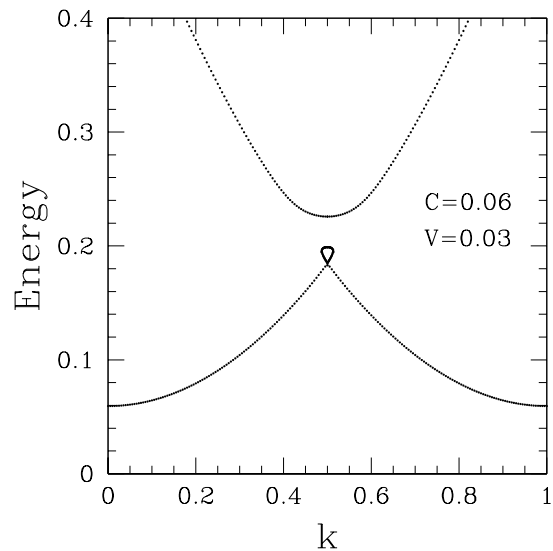}}
\caption{Two lowest Bloch bands of a BEC in an optical lattice when
$C>V$. The energy is in units of ${4\hbar^2k_L^2/m}$ and the wave number
$k$ in units of $2k_L$, where $k_L$ is the wave number of the laser light
that generates the optical lattice.}
\label{loop}
\end{figure}

One very surprising finding for the system of  BECs in optical lattices is 
the loop structure appeared in the Bloch bands as found in 
Ref.\cite{loop}(see Fig.\ref{loop}). This finding was later confirmed 
in further theoretical studies\cite{wdn,smith,smith2}. This unusual and 
unique loop 
structure has very interesting physical consequences:
First, it leads to the nonlinear Landau-Zener tunneling that is 
characteristically
different from the linear Landau-Zener tunneling\cite{lz}, in particular,
the nonzero tunneling in the adiabatic limit\cite{loop}.
Second, it destroys Bloch oscillations\cite{wn,wdn,smith}. 

However, experimental exploration of this looped band and
its related physical phenomena is yet to come forth. 
To our judgment, this experimental stalemate does not
come from the lack of experimental techniques; it is
the lack of enough theoretical guidance. On the one hand,
it is not clear what signals to look for in an experiment
to confirm the existence of the loop structure. On the other hand,
one may be concerned that the unavoidable inhomogeneity
of the BEC used in a real experiment may wash away all
the interesting physics since the predicted loop structure and its
related physics is based on  the analysis of homogeneous BECs.

The purpose of this paper is to dismiss these concerns.
Based on extensive numerical calculations, we argue that one can
confirm the existence of the loop structure in an experiment that involves
dragging a high density BEC with an accelerating optical lattice. 
Similar experiments with 
low density BECs\cite{morsch} or cold-atoms\cite{bo}
have been carried out to observe Bloch oscillations
and Landau-Zener tunneling. As we will see later, 
there are two ways to look for the signs of the loop structure: 
the destruction of Bloch oscillations and the observation of 
nonlinear Landau-Zener tunneling that shows a very distinct 
behavior from the well-known Landau-Zener tunneling.

We focus on the experimental situations similar to what is described
in Ref.\cite{morsch,josephson,bo}, where a realized BEC is 
loaded into an optical lattice and there is no trapping potential.
These experiments can be well described by the mean-field 
Gross-Pitaevskii equation in one dimension 
\begin{equation}
i\hbar {\partial \psi \over \partial t}
= - {\hbar^2 \over 2m} {\partial^2 \psi  \over \partial x^2}
  + V_{0} {\rm cos}[2k_{L}(x-{at^2\over 2})] \psi
  + { 4 \pi \hbar^2 a_s \over m} |\psi|^2 \psi ,
\label{nlse_ori}
\end{equation}
where $m$ is the atomic mass, $k_{L}$ is the wave vector of the 
laser light that generates the optical lattice, $a_s$ 
is the $s$-wave scattering length between atoms, 
$a$ is the acceleration, and $V_{0}$ is the strength of the potential 
which is proportional to the laser intensity. 
In our calculations, Gaussian functions are used to 
simulate the inhomogeneous BECs loaded in optical lattices
in the real experiments. Strictly, the lateral expansion
of the BEC has certain effects on the longitudinal motion\cite{lateral}.
In this paper we only consider the case where the lateral motion 
is negligible\cite{choi}. In experiments, another possible setup for 
quasi-one dimensional dynamics is to
confine the lateral motion\cite{1d,twzer}.

Without the acceleration, $a=0$, the system becomes a
nonlinear periodic system for which we can define
Bloch waves as for a linear periodic system, 
\begin{equation}
\psi(x)=e^{ikx}\psi_k(x)\,,
\end{equation}
where $\psi_k$ is periodic, $\psi_k(x+\pi/k_{L})=\psi_k(x)$. 
The Bloch state satisfies
\begin{eqnarray}
\mu(k)\psi_k&=& - {\hbar^2 \over 2m}\Big({\partial\over\partial x}+
ik\Big)^2\psi_k+\nonumber\\
&&+ V_{0} {\rm cos}(2k_{L}x) \psi_k
+ { 4 \pi \hbar^2 a_s \over m} |\psi_k|^2 \psi_k\,.
\label{nlse_b}
\end{eqnarray}
The eigenenergies (or more precisely, the chemical potentials) $\mu(k)$ then 
form Bloch bands in the Brillouin zone. As shown in Fig.\ref{loop},
when the interaction between atoms gets larger than a
certain critical value, a loop structure is formed in the 
Bloch bands.

In the following, we first briefly describe our
numerical methods. Then, we present our numerical results
on nonlinear Landau-Zener tunneling and
Bloch oscillations, and explain how to look for signs of the loop 
structure through these two phenomena. In the end, we discuss the
relevance of our results to the experiments.

\section{Numerical method}
For the convenience of numerical calculations, we cast Eq.(\ref{nlse_ori})
into a dimensionless form
\begin{equation}
i {\partial \phi \over \partial t}
= - {1 \over 2}{\partial^2 \phi \over \partial x^2}
  + V{\rm cos}(x-{1\over 2}at^2) \phi
  + C |\phi|^2 \phi \,,\\
\label{nlse_comp}
\end{equation}
where we have used the following set of scaled variables, 
\begin{eqnarray}
\tilde{x} = 2 k_{L} x\,,&
\tilde{t} = {4\hbar k_{L}^2 \over m} t\,,& 
\tilde{a}={m^2\over 8\hbar^2k_L^3}a\,,\nonumber\\
\phi = { \psi \over \sqrt{n_{0}} }\,,&
V = {m \over 4\hbar^2 k_{L}^2 } V_{0}\,,& 
C = { \pi n_{0} a_s \over k_{L}^2 }\,,\nonumber
\end{eqnarray}
with $n_{0}$ being the peak density of the BEC cloud.
In writing down Eq.(\ref{nlse_comp}), we dropped the tildes 
(replacing $\tilde{x}$ by $x$, etc.) without causing confusion.
We use the Crank-Nicholson method for the numerical solution of 
Eq. (\ref{nlse_comp}). This method preserves the unitarity of 
the time-evolution, and yields good convergence of the solutions 
for moderate values of the coupling strength $C$. Note that for
the experiment where the lateral motion is confined, $C$ has
different definitions, see one example in Ref.\cite{twzer}.

In many experiments\cite{burger,morsch}, a BEC cloud has a  typical 
size of order 10 $\mu$m, covering 100-200 wells of an optical lattice.
To model such inhomogeneous BEC clouds, we use
a Gaussian wave packet as the initial state and then turn it
into a inhomogeneous BEC Bloch wave by adiabatically turning 
on the optical lattice.

The lattice strength $V$ is taken to be smaller than 0.4 (or 3.2
in units of the recoil energy $\hbar^2 k_L^2/2m$). 
This choice serves two purposes. First, it guarantees that there be 
only one bound state inside each well so that we can
use the separation of a BEC cloud to measure the tunneling probability
as we will explain below. Second, it means that the mean field 
theory (\ref{nlse_ori}) is a good description of the BEC system.  

\section{Nonlinear Landau-Zener Tunneling}
The nonlinear Landau-Zener tunneling has been studied quite extensively
in Ref.\cite{loop}, where it is found to be very different from
the linear Landau-Zener tunneling\cite{lz}. In particular, when $C>V$, 
the tunneling probability no longer depends on the sweeping rate
exponentially (sweeping rate is the acceleration for the system of
a BEC in an accelerated lattice).
Moreover, in the adiabatic limit where the acceleration approaches zero,
the tunneling probability approaches a finite value, instead of
zero as in the linear case. This non-zero tunneling probability in
the adiabatic limit is the direct result of the loop structure:
when a Bloch state is driven to the edge of the loop, it has to
split, resulting in tunneling \cite{loop}. Therefore, experimental observation
of this adiabatic tunneling, along with the non-exponential dependence
of tunneling probability on the acceleration, can be viewed 
as a direct evidence of the looped band.
 
However, the analysis in Ref.\cite{loop} is based on a simplified
two-level model derived with the assumption of homogeneity of BEC. 
This may leave experimentalists wonder whether the unique characteristics 
of nonlinear Landau-Zener tunneling can be observed in real experiments, 
where BECs are inhomogeneous and span only a finite range of space. 
This concern is legitimate, but our numerical results show that the
inhomogeneity does not blur up the essential physics.

\begin{figure}[!htb]
\centerline{\includegraphics[width=8.25cm]{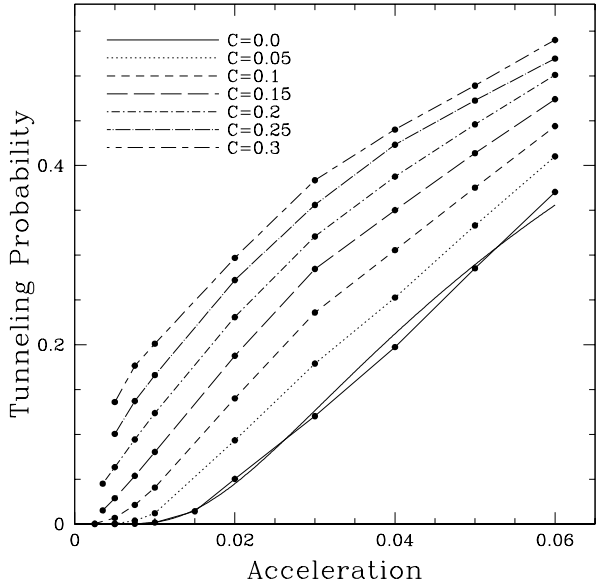}}
\caption{Tunneling probability as a function of acceleration $a$ for
various values of $C$ and $V=0.2$. The case of $C=0$ is compared to the
exponential function $e^{-0.061/a}$ indicated by solid line without 
($\bullet$).
}
\label{tunprob}
\end{figure}
In our numerical simulations, we calculate the tunneling probability 
through the separation of a BEC cloud. As we mentioned earlier,
the lattice strength is chosen such that there is a only one band
below the well barrier. As a result, the part of a wave packet tunneled into 
the upper band will not be dragged along the lattice while the
part remained in the lower band will be dragged along. This leads to
a separation of a BEC cloud after a certain time of acceleration. 
By integrating the left-behind wave packet, we obtain the tunneling
probability. This technique was actually used in experiments to
measure the tunneling probability \cite{morsch,bo}.

Fig. \ref{tunprob} shows our results of the nonlinear Landau-Zener
tunneling probability with $V=0.2$ for various values of $C$. 
The initial Gaussian wave packet is
$\phi(x,t=0) = e^{- x^2 / \omega^2}$, where $\omega$ is the  
width of the condensate and its typical values used is $\omega=325$.
Optical lattice potential is turned on from $V(x)=0$ to $V(x)=V$ 
for $t \le 40$ to achieve an inhomogeneous BEC Bloch wave. Then
it is boosted with an acceleration $a$ for two Bloch periods,
and moves with a constant velocity afterward.

In general, tunneling probability is greater for larger $C$ at a fixed 
acceleration $a$, and greater for large accelerations $a$ at a fixed $C$.
For $C<V$, the tunneling probability changes with acceleration
$a$ rather exponentially and in particular, it goes to zero 
when $a \rightarrow 0$.
This is exactly what is predicted in Ref.\cite{loop}. 
Therefore, this result implies that the inhomogeneity has 
only negligible impact on the tunneling probability when $C<V$.

When $C>V$, we also do not see much impact of the inhomogeneity of BEC:
the tunneling probability does not depend on the acceleration $a$ 
exponentially and in particular, when the data is 
extrapolated to the adiabatic limit, $a \rightarrow 0$, the tunneling
probability seemingly tends to a non-zero value. This clearly shows that
these two unique characteristics of nonlinear Landau-Zener
tunneling are not washed away
by the inhomogeneity  of the BEC cloud. The experimental observation
of them will be viewed as the confirmation of the loop structure.
However, since the acceleration can not be made arbitrarily small
in a real experiment and our simulations, one may argue the validity and
confidence on the extrapolation of the data to the adiabatic limit of zero
acceleration. We believe that this problem can be partially 
solved by repeating the experiment with many different densities of 
BECs. Consistency among different sets of data will be a strong
support of the validity of extrapolation. On the other hand,
the observation of non-exponential dependence of the tunneling probability
on the acceleration will provide an unambiguous evidence. 

\section{Bloch Oscillations}
Bloch oscillations occur when a Bloch state is driven
across the Brillouin zone by a small external field\cite{bo}. 
For a BEC in an optical lattice, Bloch oscillations
can be achieved by dragging the lattice with a small
acceleration as reported experimentally 
in Refs.\cite{morsch,josephson}.
In order to observe these oscillations, a key requirement
is that the acceleration must be small enough so that the tunneling 
to the upper band is negligible. If tunneling probability 
into the upper band is increased, say, by large accelerations, 
Bloch oscillations can be destroyed. 

What is interesting with BECs is that we have another
way to increase the tunneling probability besides
increasing the acceleration. It is to increase 
the density of the BEC. Especially, as discussed
in the last section, when the density is high
enough such that $C>V$, there is a non-zero lower limit
on the tunneling probability as the result of the loop
structure in the energy band. In other words,
no matter how small the acceleration is, we will
not be able to observe Bloch oscillations due to the non-zero adiabatic 
tunneling when $C$ is big enough.
Therefore, the observation of breakdown of Bloch oscillations
will provide another way to detect the loop structure. 
\begin{figure}[!htb]
\centerline{\includegraphics[width=8.25cm]{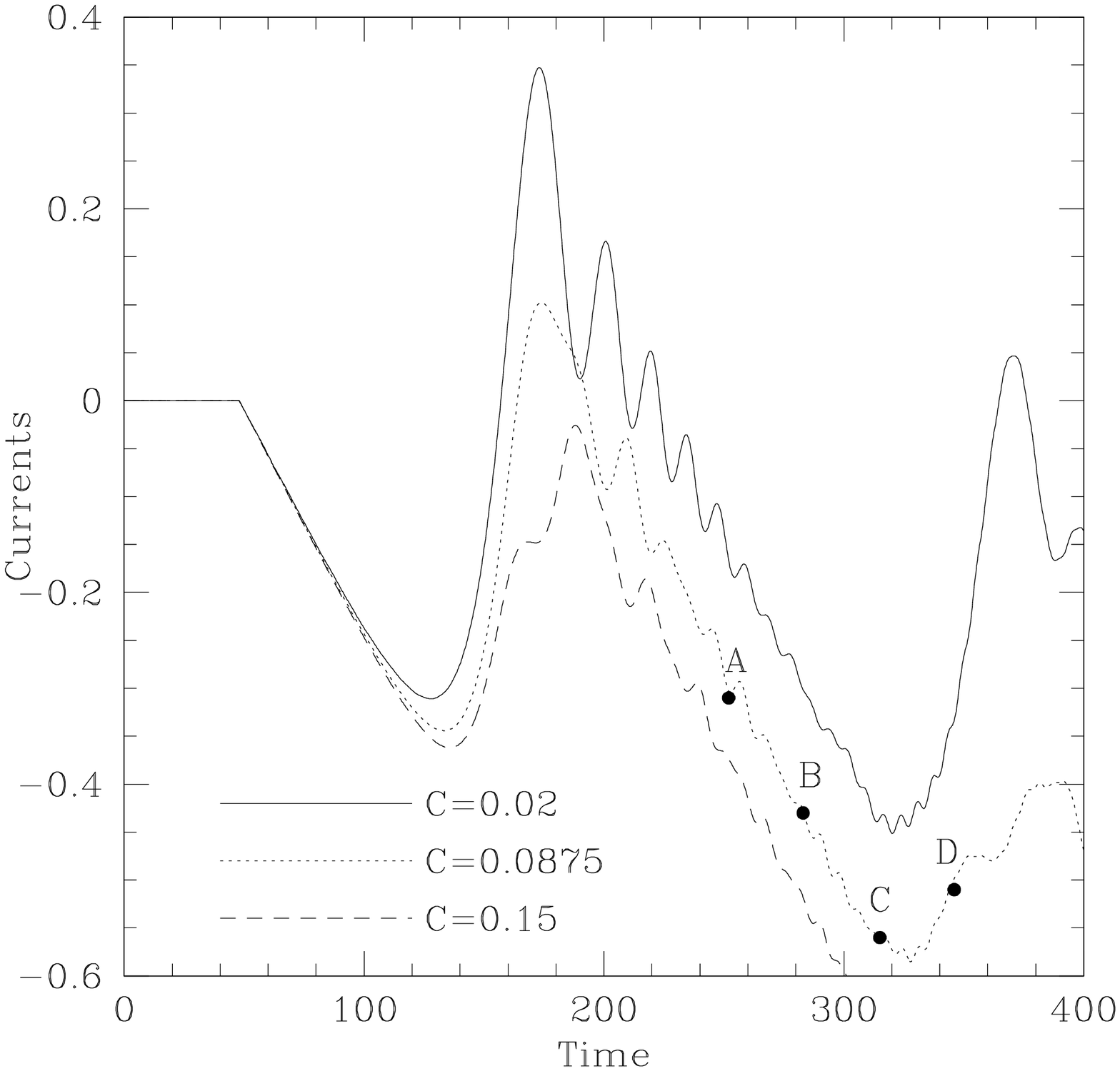}}
\centerline{\includegraphics[width=9.5cm]{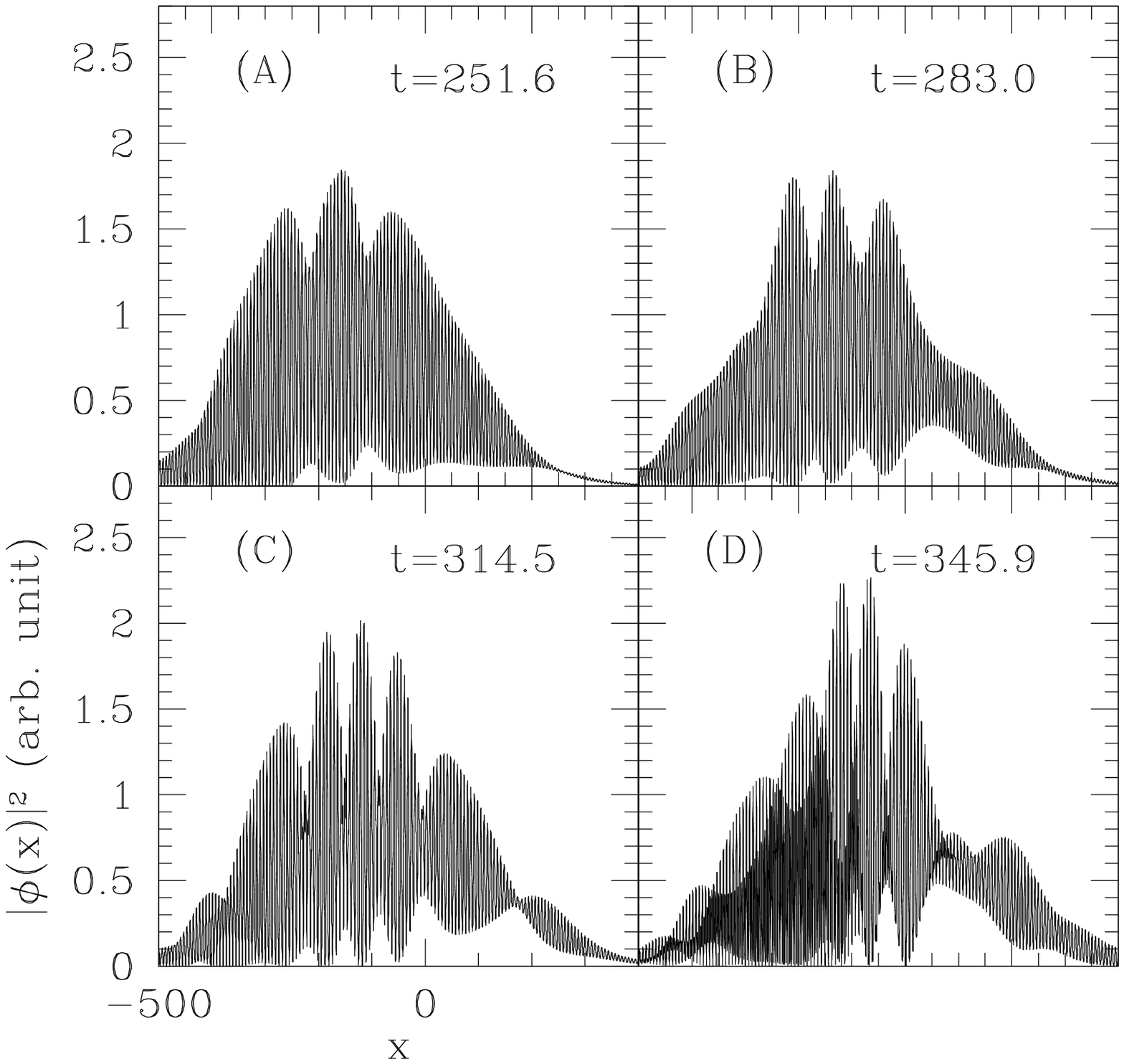}}
\caption{Top panel: Currents as a function of time for
various values of $C$ at $V=0.0875$ and $a=0.005$. Bloch oscillations
are destroyed clearly when $C>V$. Results are shown for $C$=0.02, 0.0875,
and 0.15 in the accelerating frame.
Bottom panel: The wave functions at different points of breakdown of Bloch 
oscillations, as indicated in the top panel at $t=251.6 (A), 283.0 (B), 314.5 
(C),$ and $345.9 (D)$.
The smoothness of the wave functions 
shows no signs of dynamical instability, implying the destruction of
Bloch oscillations is the result of nonlinear Landau-Zener tunneling.
}
\label{bo}
\end{figure}

In an experiment, one can repeat the measurements
with increasing densities of BECs for a fixed small acceleration.
One expects to observe Bloch oscillations when the density 
is low; as the density increases, the oscillations will
deteriorate and eventually be destroyed. This is indeed
the case as shown in Fig. \ref{bo} from our simulations.
In this figure, we show currents of the condensate,
$j = \int ({ \hbar \over m}) {\rm Im}( \phi^* {d \phi \over dx} ) dx$, 
as a function of time in the accelerating frame for a small $a$.
For $C<V$, Bloch oscillations are preserved
during the first Bloch period; however, for $C>V$, Bloch oscillations are 
disrupted and completely destroyed during the first Bloch period.
Even for the case of $C=V$, Bloch oscillations are seriously disrupted. 
\begin{figure}[!htb]
\centerline{\includegraphics[width=8.25cm]{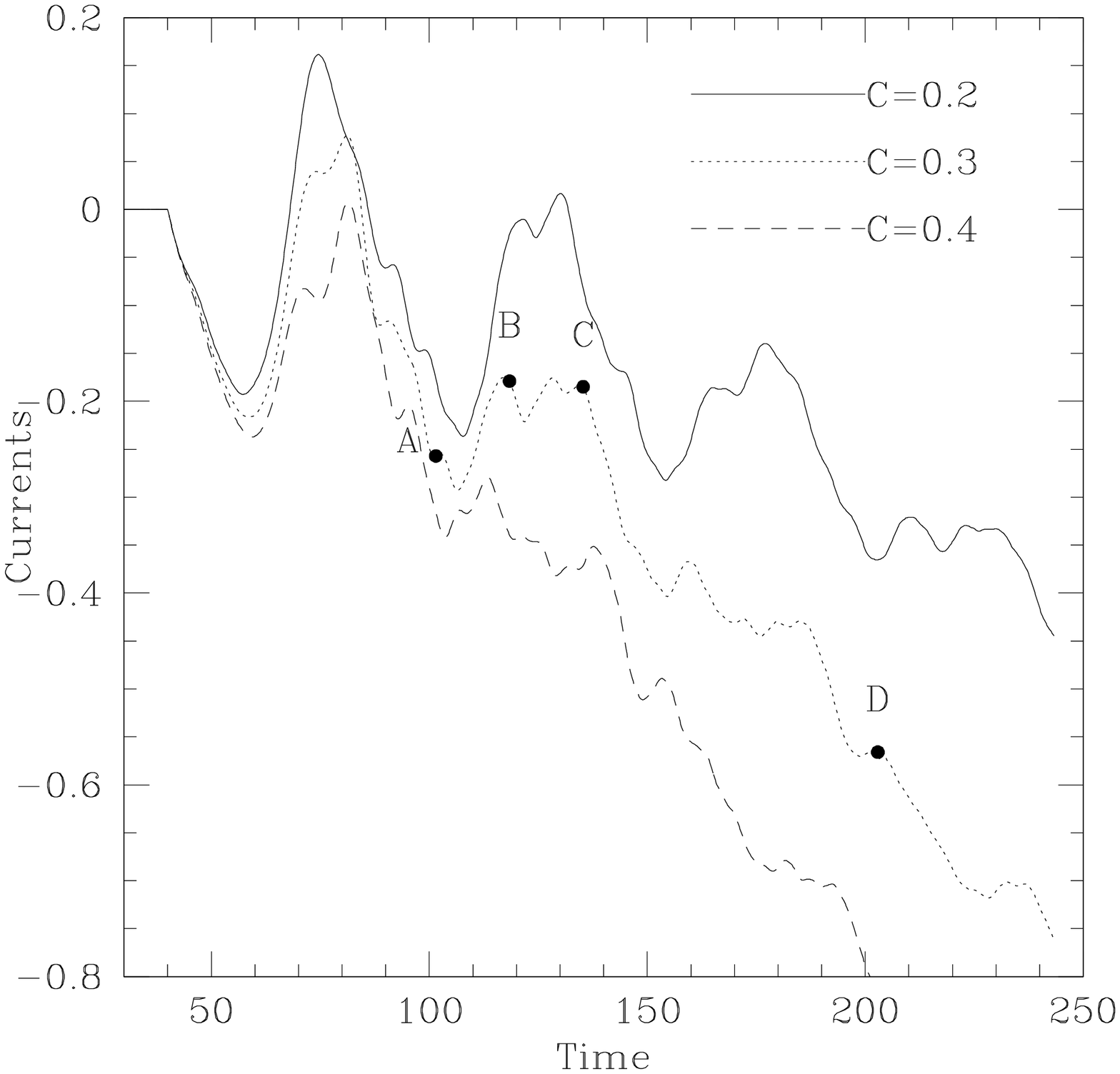}}
\centerline{\includegraphics[width=9.5cm]{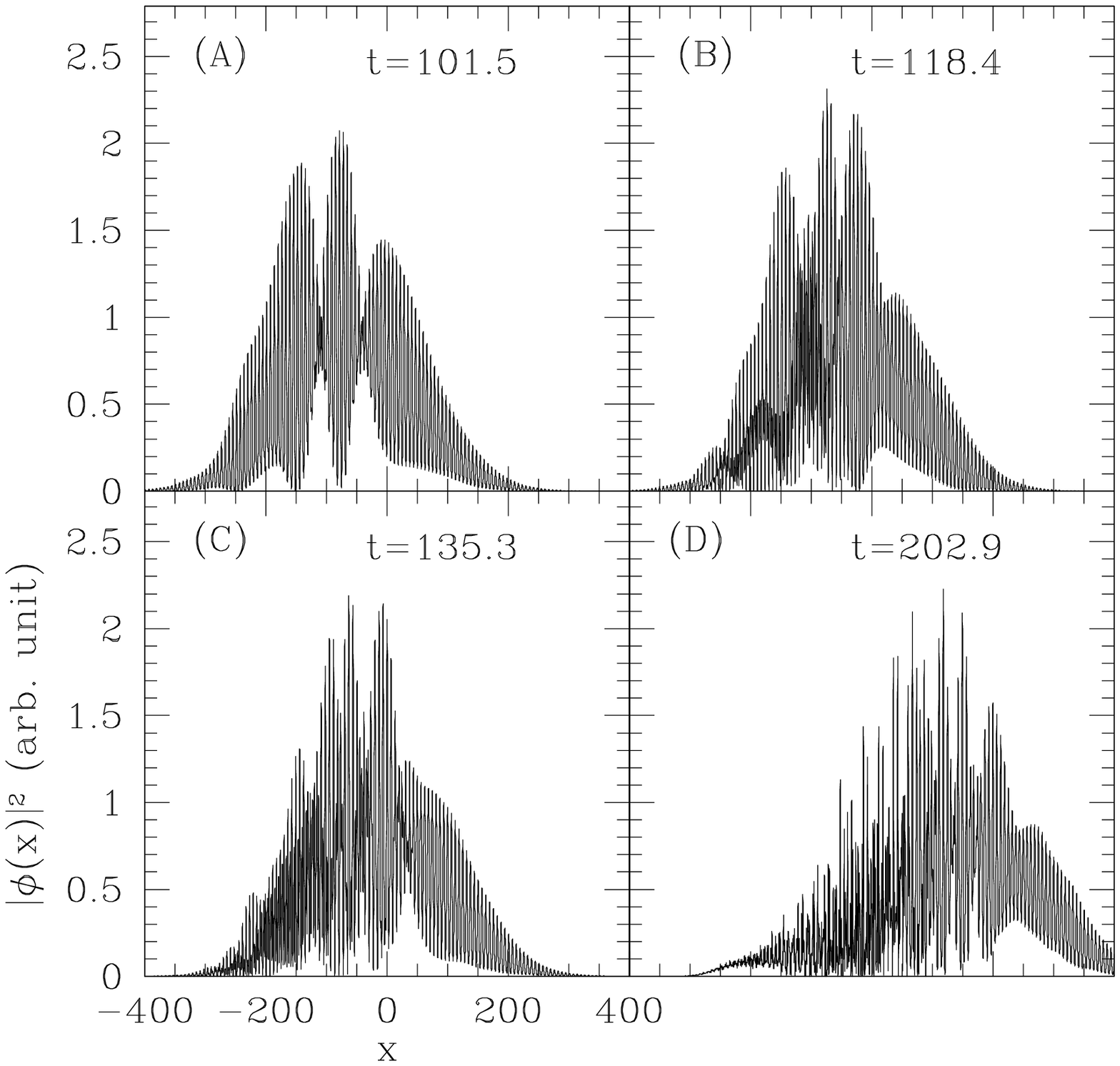}}
\caption {Top panel: Currents as a function of time
with $V$ = 0.3175 and $a$ = 0.02. Results are shown for $C = 0.2, 0.3$, and 
$0.4$ in
the accelerating frame, and we see the destruction of 
Bloch oscillations even when $C<V$, indicating that it is
caused by the dynamical instability.
Bottom panel: Densities of the wave function, $|\phi(x)|^2$, for $C=0.3$  
at $t=101.5 (A), 118.4 (B), 135.3 (C),$ and $202.9 (D)$. The ``spiky'' or 
``messy'' wave functions at points $C$ and $D$ are the result of
dynamical instability. }
\label{di_c0p3_crnt}
\end{figure}

In our view, this is a much better way to detect the loop 
structure shown in Fig.\ref{loop} than the method discussed 
in the last section regarding the nonlinear Landau-Zener tunneling.
With increasing densities, the measurement of the tunneling
probability based on the separation of a BEC cloud will become
more difficult because the higher the density the faster the cloud expands.
This makes the separation more difficult. With Bloch
oscillations, we do not need to worry about this difficulty.

One caution we need to take is that there is another
mechanism for breakdown of Bloch oscillations, the dynamical
instability of Bloch waves discussed in Ref.\cite{wn,wdn,smith2}. The 
instability
will cause the system stray away from the Bloch states with very small
amount of perturbations or noises. This is demonstrated in 
Fig.\ref{di_c0p3_crnt}, where we see the destruction of Bloch oscillations
even when $C<V$ for a very small acceleration. The ``spiky'' or ``messy''
wavefunctions in the bottom panel signal the onset of dynamical instability.

The dynamical instability always exists in the neighborhood of $C\sim 
V$\cite{wdn,smith2}, 
where the breakdown of Bloch oscillations starts to take place. 
As a result, one must make sure that the breakdown of Bloch oscillations 
is the result of the loop structure, instead of the dynamical 
instability. One way is to use weak optical lattices (small values of V).
In this case, the growth rates of unstable modes
are small therefore the dynamical instability will not dominate
in the first few oscillations as seen in the bottom panel of Fig.\ref{bo}. 
These smooth wave functions indicate that the dynamical instability
is yet to come into play. Therefore, the destruction of Bloch oscillations
in the upper panel is purely due to the nonzero adiabatic tunneling
resulted from the loop structure.\\

\section{Conclusion}
In summary, using extensive and realistic calculations 
we have demonstrated experimental feasibility to detect 
the loop structure in the BEC Bloch bands.
We suggested two possible scenarios: the observation of breakdown of Bloch 
oscillations
and nonlinear Landau-Zener tunneling in an accelerating lattice.
Experiments similar to what we suggest have already been carried out with 
low density BECs\cite{morsch}, where the corresponding coupling strength 
is in the range of $C=0.026 \sim 0.04$. 
The values used in our calculations cover the range of $C = 0.05 \sim 0.3$.
This means that the signatures of loop structure studied in this paper
can be observed by increasing the density of BECs by two to ten times. 
This is certainly possible with the current experimental set-ups.
The density of the BEC in Ref.\cite{morsch}, $10^{14}$cm$^{-3}$,
can be increased to meet the requirement considering 
the BECs of density as high as $3\times 10^{15}$cm$^{-3}$
have already been achieved\cite{high}.

\begin{acknowledgments}
Wu was supported by the NSF and the LDRD of ORNL, managed by UT-Battelle for 
the USDOE (DE-AC05-00OR22725).
\end{acknowledgments}


\begin{thebibliography}{99}
\bibitem{choi}D.-I. Choi and Q. Niu, Phys. Rev. Lett. 
{\bf 82}, 2022 (1999).
%
\bibitem{wn}B. Wu and Q. Niu, Phys. Rev. A {\bf 64}, 
061603 (2001).
%
\bibitem{zobay}O. Zobay and B.M. Garraway, Phys. Rev. A {\bf 61},
033603 (2000).
%
\bibitem{burger}S. Burger, F.S. Cataliotti, C. Fort,
F. Minadri, and M. Inguscio, M.L. Chiofalo, and M.P. Tosi,
Phys. Rev. Lett. {\bf 86}, 4447 (2001).
%
\bibitem{morsch}O. Morsch, J.H. M\"uller, M. Cristiani,
D. Ciampini, and E. Arimondo, Phys. Rev. Lett. {\bf 87},
140402 (2001); M. Cristiani, O. Morsch, J.H. M\"uller,
D. Ciampini, and E. Arimondo, Phys. Rev. A {\bf 66}, 043409 (2002). 
%
\bibitem{wdn}B. Wu, Roberto B. Diener, and Q. Niu, 
Phys. Rev. A {\bf 65}, 025601 (2002).

\bibitem{josephson}B.P. Anderson and M.A. Kasevich,
Science {\bf 282}, 1686 (1998).

\bibitem{squeeze}C. Orzel, A.K. Tuchman, M.L. Fenselau, M. Yasuda,
and M.A. Kasevich, Science {\bf 291}, 2386 (2001).

\bibitem{mott}M. Greiner, O. Mandel, T. Esslinger, 
T.W. H\"{a}nsch, and I. Bloch, Nature {\bf 415},
39 (2002); D. Jaksch, C. Bruder, J.I. Cirac, C.W. Gardiner,
and P. Zoller, Phys. Rev. Lett. {\bf 81}, 3108 (1998);
V.A. Kashurnikov, N.V. Prokof'ov, and B.V. Svistunov, 
cond-mat/0205510.

\bibitem{loop}B. Wu and Q. Niu, Phys. Rev. A {\bf 61},
023402 (2000).


\bibitem{smith}D. Diakonov, L.M. Jensen, C.J. Pethick, 
H. Smith, Phys. Rev. A {\bf 66}, 013604 (2002).

\bibitem{smith2}M. Machholm, C.J. Pethick, and H. Smith, 
Phys. Rev. A {\bf 67}, 053613 (2003).

\bibitem{lz}L.D. Landau, Phys. Z. Sov. {\bf 2}, 46 (1932);
G. Zener, Proc. R. Soc. London, Ser. A {\bf 137}, 696 (1932).

\bibitem{bo}M. Ben Dahan, E. Peik, J. Reichel, Y. Castin,
and C. Salomon, Phys. Rev. Lett. {\bf 76}, 4508 (1996);
M. Raizen, C. Salomon, and Q. Niu, 
Physics Today July 1997, p.30.

\bibitem{lateral}P. Massignan and M. Modugno, 
Phys. Rev. A {\bf 67}, 023614 (2003).

\bibitem{1d}A. G\"orlitz, J.M. Vogels, A.E. Leanhardt,
C. Raman, T.L. Gustavson, J.R. Abo-Shaeer, A.P. Chikkatur,
S. Gupta, S. Inouye, T. Rosenband, and W. Ketterle,
Phys. Rev. Lett. {\bf 87}, 130402 (2001).

\bibitem{twzer}R.B. Diener, B. Wu, M.G. Raizen, and Q. Niu,
Phys. Rev. Lett. {\bf 89}, 070401 (2002).


\bibitem{high}D.M. Stamper-Kurn, M.R. Andrews, A.P. Chikkatur, 
S. Inouye, H.-J. Miesner, J. Stenger, and W. Ketterle,
Phys. Rev. Lett. {\bf 80}, 2027 (1998).
\end{thebibliography}
\end{document}